\documentclass{article}

\usepackage{PRIMEarxiv}

\usepackage[utf8]{inputenc} % allow utf-8 input
\usepackage[T1]{fontenc}    % use 8-bit T1 fonts
\usepackage{hyperref}       % hyperlinks
\usepackage{url}            % simple URL typesetting
\usepackage{booktabs}       % professional-quality tables
\usepackage{amsfonts}       % blackboard math symbols
\usepackage{nicefrac}       % compact symbols for 1/2, etc.
\usepackage{microtype}      % microtypography
\usepackage{lipsum}
\usepackage{fancyhdr}       % header
\usepackage{graphicx}       % graphics
\graphicspath{{media/}}     % organize your images and other figures under media/ folder

\title{Is Decentralized AI Safer?
}

\author{
  Casey Clifton \\
  Algovera \\
  \texttt{caseyclifton@proton.me} \\
   \And
  Richard Blythman \\
  Algovera \\
  \texttt{richardblythman@gmail.com} \\
   \And
  Kartika Tulusan \\
  Algovera \\
  \texttt{ktulusan@gmail.com} \\
}

\begin{document}
\maketitle

\begin{abstract}
Artificial Intelligence (AI) has the potential to significantly benefit or harm humanity. At present, a few for-profit companies largely control the development and use of this technology, and therefore determine its outcomes. In an effort to diversify and democratize work on AI, various groups are building open AI systems, investigating their risks, and discussing their ethics. In this paper, we demonstrate how blockchain technology can facilitate and formalize these efforts. Concretely, we analyze multiple use-cases for blockchain in AI research and development, including decentralized governance, the creation of immutable audit trails, and access to more diverse and representative datasets. We argue that decentralizing AI can help mitigate AI risks and ethical concerns, while also introducing new issues that should be considered in future work.
\end{abstract}

% keywords can be removed

\section{Introduction}

\subsection{Artificial Narrow Intelligence}
Over the past decade, progress in Artificial Intelligence (AI) has been rapid, exceeding the expectations of expert AI researchers and developers \cite{steinhardt_2022_ai}. As a result, applications using specialist, ‘narrow’ models now permeate many aspects of modern life \cite{page_2018_the}. These algorithms power internet search engines and smart home devices, control online advertising, run safety systems in cars, assist the writing of code, emails and essays, play games, and can even produce original art.

Despite receiving \$93.5 billion of private investment in 2021 \cite{zhang_maslej_brynjolfsson_etchemendy_2022}, and powering the tech-giants that dominate global consumer markets, narrow AI systems have many well-documented safety and ethical concerns \cite{tegmark_2015_benefits}. For example, due to biased training data, commercial AI systems can fail to perform equitably once deployed in real-world applications, such as in the case of gender discrimination by facial recognition systems \cite{Buolamwini2018GenderSI}. Furthermore, in the recent surge of AI generated text and visual content by models such as GPT-3 \cite{Brown2020Language} and DALLE-2 \cite{Ramesh2022Hierarchical}, ethical and safety concerns arise regarding the propagation of misinformation \cite{kreps_mccain_brundage_2022}.

\subsection{Artificial General Intelligence}
Many investors, developers and researchers look beyond today’s narrow AI and toward Artificial General Intelligence (AGI): systems that are at least as smart as humans, and capable of inventing original and creative solutions to the world's most pressing problems \cite{goertzel_2013_artificial}. Experts vary widely on when they predict AGI will be achieved, but almost all believe it’s possible \cite{forecasting, clarke_2021_survey}. While the potential benefits of AGI dwarf those of narrow AI, so do its risks \cite{mclean_2021_the}.

Since Nick Bostrom published his book on the topic, ‘Superintelligence’, in 2014, the potential dangers of AGI have received increased attention from the international AI community \cite{bostrom2014}. The core challenge of AGI is the alignment problem; namely, how can we ensure that an AGI will be developed in-line with human preferences, and who decides what preferences are prioritized \cite{yudkowsky_2016_ai}. While evidence and concern for these issues is growing, only approximately 300 people are working full-time on AI safety \cite{hilton_2022_preventing}. By contrast, billions are invested annually into creating new and more powerful AI \cite{zhang_maslej_brynjolfsson_etchemendy_2022}.

The development of AI models today is led by a handful of large technology companies, who effectively have monopolistic control over the direction of this technology. For example, Microsoft invested \$1 billion in OpenAI, who only had 150 employees at the time, while an even smaller team controlled and managed the development of their most advanced language model, GPT-3 \cite{siddarth_2021_how}. The centralization of AI research and development (R\&D) risks the creation of AGI in-line with a narrow and limited set of moral-views that are not representative of all those who will be affected by it \cite{discriminating}.

\subsection{Decentralization}

Blockchain technology, pioneered by the launch of Bitcoin in 2008 \cite{nakamoto_2008_bitcoin}, has received significant attention due to financial speculation on cryptocurrencies. However, it has many other applications. Notable examples include work on public goods funding \cite{buterin_2019_a}, charitable donations \cite{8946019}, carbon credit markets \cite{8486626}, and the decentralization of scientific research \cite{lehner_2017_funding}.

Blockchain also has use-cases in AI. For example, it can facilitate and incentivize the secure exchange of personal data that improves AI model performance \cite{computetodata}. It also provides an immutable record for auditing \cite{falco}, new funding mechanisms \cite{thomason_2018_blockchainpowering}, and a means for community-driven governance \cite{montes_2019_distributed}. 

We use the umbrella term ‘decentralized AI’ to capture all of these diverse use-cases, and in this paper, focus on those that affect AI safety. In Section 2, we demonstrate how decentralized AI can mitigate risks and promote the ethical use of AI. In Section 3, we discuss the risks of decentralization that must be considered in future work.

\section{Benefits of Decentralized AI}
Decentralized AI is still in its infancy, yet it can already mitigate a number of AI risks. In this section, we address how decentralizing governance and funding aligns AI R\&D with more representative perspectives and moral views, and how decentralizing data ownership and exchange can reduce model bias and improve technical solutions.

\subsection{Safer Governance of Research and Development}

To identify safe and ethical frameworks, scientific communities often seek a diverse range of perspectives by hosting open and collaborative discussions. In the case of gene-editing, for example, when a doctor claimed to have successfully edited a human genome for the first time, the scientific community responded by launching an international commission to develop new ethical guidelines \cite{a2019_new}. Indeed, there are now international conferences on AI safety with similar purposes \cite{a2015_ai}. 

Building upon this open and collaborative approach, Siddarth et al. argue that a pluralistic and participatory model, rather than a singular centralized one, will yield safer and more ethical AI \cite{siddarth_2021_how}. Blockchain facilitates this vision of AI by helping to coordinate and incentivize participation through Decentralized Autonomous Organizations (DAOs) \cite{hassan_2021_decentralized}. In a DAO, all members can contribute to enforceable collective decisions about what AI projects get funded, what their terms of service are, and who can use them.

In 2018, a group of 4000 Google employees petitioned that their technology should not be used to advance autonomous warfare, and were able to convince the company’s management to forgo a particular military contract \cite{wakabayashi_2018_google}. In this scenario, the company executive’s profit-driven incentives were not aligned with the ethical view of a larger and more representative group. However, because the employees do not hold real voting power, this decision can be overturned at any time. Furthermore, it has not led to a strict company-wide policy; despite the successful petition, Google still supports military efforts through its ads platform. By contrast, a DAO governed by a diverse group of voters could implement ethical and representative policies that are prioritized in every business decision \cite{montes_2019_distributed}.

In current community-driven efforts to develop AI, we see further evidence of prioritizing ethical decision-making through open discussions. Eleuther AI, for example, is a “grassroots collective of researchers working to open source AI research” \cite{eleuther}. In the process of training and open sourcing a large language model similar to GPT-3, they had extensive and public discussions on ethical and safety concerns \cite{leahy_2021_why}. In another example, the Distributed Artificial Intelligence Research institute was formed explicitly to “create space for independent, community-rooted AI research free from Big Tech’s pervasive influence” \cite{a2021_timnit}. The community-driven decision-making process and ethical priorities of these groups demonstrate the value of diversifying AI efforts, rather than containing them within a handful of large technology companies. While governance of these two groups is not conducted via a blockchain, doing so could secure their voting systems and promote further decentralization.

In fact, early examples of governing AI R\&D using blockchain already exist. Algovera and SingularityNET, who leverage blockchain to enable decentralized decision-making, fund work on beneficial and ethical AI, focusing on topics such as climate change, food security, and precision medicine \cite{algoverse, research}. These groups demonstrate that blockchains offer alternative pathways for individuals or groups to acquire funding, and can help new businesses and non-profit organizations create ledgers, treasuries and voting systems \cite{thomason_2018_blockchainpowering}. Blockchains can therefore formalize independent and distributed work on AI, ensuring that diverse perspectives are incorporated into enforceable governance frameworks.

\subsection{User-aligned AI}
In recent years, human-centered AI has emerged as a core principle for developing safer and more ethical AI \cite{humanc}.  Bondi et al., for example, design a model that prioritizes the views of communities who will be affected by AI to more equitably distribute the social good it can do \cite{bondi_xu_acosta}. Blockchains facilitate such approaches, empowering users and communities to guide AI development through ownership and governance stakes in the platforms they use \cite{nabben_2021_grounding}. A user-ownership model could also bring relief to the majority of people who are concerned about how their data is currently used by companies and governments \cite{hori_2021_selfsovereign}. 

\subsection{AI Alignment via Representative Feedback}
In a decentralized setting, AI models can learn from more diverse and representative human feedback. Techniques such as Imitative Generalization and Recursive Reward Modelling use a human-in-the-loop model to maintain alignment between the desired and realized outcomes of an agent's actions \cite{barnes_2021_imitative, leike_2018_scalable}. The resulting agent, therefore, will tend towards alignment with the particular individuals guiding it. In an open, decentralized setting, diverse groups can more easily coordinate and work together, and therefore train AI that is aligned with a greater plurality of ethical views and perspectives.

\subsection{Immutable Audit Trails}
At its core, a blockchain is an immutable public ledger, meaning records stored on it are transparent, cannot go missing, and cannot be manipulated (provided the network validating the blockchain is strong enough) \cite{8805074}. Auditing a decentralized AI system, which uses a blockchain to record database updates, AI model retraining, and code changes, may be easier than when AI systems are developed and deployed behind closed doors \cite{8399460}. Improved auditing allows for more informed regulation and monitoring of AI system capabilities and biases. Moreover, blockchains can be used as immutable and public records of accidents and mistakes by AI systems. In other domains where the safety of complex systems is critical, such as aviation, such data sharing plays a key role in improving safety over time \cite{Syed2016Black}.

\subsection{Data Sovereignty}
Blockchain technology enables individuals to own and manage access to their own data \cite{hori_2021_selfsovereign}. For example, Ocean protocol has created a marketplace where users can securely publish their data for researchers to train AI models on \cite{oceanmarket}. With this data sovereignty comes a number of benefits. Firstly, rather than sacrificing or risking privacy without consent or financial compensation, individuals can be directly rewarded as their data adds value to AI applications that they (or a delegated data trust) has consented to \cite{nabben_2021_decentralised}. This mechanism helps to redistribute wealth created by AI systems. Secondly, it significantly reduces the risk of major privacy breaches, because no single entity controls access to sensitive data \cite{mhle_2018_a}. In the Cambridge Analytica case, for example, Facebook supplied user data to a foreign company for analytics used in political campaign strategies \cite{chan_2019_cambridge}. Had access to this data been controlled by blockchain-based mechanisms, there would be no single point of failure, as data could only be accessed with the individual private keys of each owner.

The impacts of data sovereignty also extend to expert annotators, who determine the ‘ground truth’ labels for data that supervised machine learning models train on. In centralized settings, annotators are rewarded by one-off payments to help create systems that aim to automate their jobs. By contrast, decentralized AI systems enable annotators to have an ownership stake in the data they create, and can therefore receive some of the value generated by the AI they are training \cite{blythman_2021_using}.

\subsection{Data Availability}
With ownership over their own data, and a means to securely monetize it, individuals become incentivized to contribute to new datasets larger and more diverse than their centralized counterparts \cite{shen_2020_blockchainbased}. As a result, developers can train and test models on more representative datasets before they are deployed in real world applications. Indeed, prior research explores the benefits of securely training AI on siloed data through the use of federated learning \cite{li_2020_federated}. Blockchain extends such approaches by incentivizing data exchange through monetization and improving the security of data pipelines \cite{8843900}. In addition, with access to new data, models may emerge that specialize on niche problems or small subsets of the population that are not sufficiently represented in existing datasets.

Data that is owned by individuals, rather than by centralized entities, can also be distributed more widely. This allows independent researchers and developers with unique and diverse agendas to work on new applications of AI. For example, rather than one hospital controlling access to patient data, and only licensing it to a small number of researchers, patients can monetize it via a blockchain. Researchers from around the world can then access that data and use it to perform a wider variety of studies \cite{Chen2019BlockchainIH}. Crucially, this work can leverage techniques such as `compute-to-data' to ensure confidential information remains private and secure \cite{computetodata}.

New, richer datasets may also assist efforts to solve the alignment problem. Traditionally, aligning AI models with human preferences is difficult, because it is challenging to accurately express preferences without loopholes or misinterpretations. This leads to cases of ‘reward-hacking’ and ‘negative side effects’, where an AI agent performs unintended or suboptimal actions\cite{Amodei2016ConcretePI}. For example, when designing a cleaning robot one might use its visual sensors to define the reward function: "once you can no longer see any mess, you will get a reward". Even in this toy example, however, an agent may find shortcuts, such as covering its eyes, or continually re-making the mess only to clean it up again. Stuart Russell proposes three new design principles to solve this: the machine’s only objective is to maximize its realization of human preferences, the machine is initially uncertain about what those preferences are, and the ultimate source of information about human preferences is human behavior \cite{russell_2020_human}. The Inverse Reinforcement Learning (IRL) paradigm, in which an AI’s goal is to identify the reward function of the humans it’s observing, puts these principles into practice \cite{irl}. Decentralization benefits IRL agents because it offers access to private data, as well as transparency on financial transactions and behaviors, which together yield richer insights on individuals’ preferences. Agents able to observe such data are thus more likely to accurately infer complex and dynamic human preferences.

\section{Risks of Decentralized AI}
While decentralization has the potential to mitigate risks from AI, it also introduces new ones. It is critical that we discuss these to present an even assessment of the technology. We also recognize the risks of centralized AI that decentralization is not immune to, such as poor model transparency and explainability \cite{eitelporter_2020_beyond}. While outside the scope of this paper, these should remain a focus for centralized and decentralized AI work going forward.

\subsection{Vulnerabilities of Decentralized Governance}
It is not clear that fully decentralized governance is a perfect model for solving ethical and safety concerns. For example, a coordinated group of bad actors or ‘bots’ could acquire sufficient governance tokens to effectively take control of a DAO and its treasury in a ‘Sybil attack’ \cite{douceur_2002_the}. Moreover, because accounts (`wallets') on a blockchain are anonymous cryptographic key-pairs, it is relatively straightforward for one person to control many wallets while maintaining the illusion of distributed ownership and control.

A decentralized governance structure enables voting at scale. This is good for diversity, however it increases the probability of leaking sensitive or dangerous information. For example, consider a DAO that has discovered techniques that could plausibly lead to AGI, and must share this information with voters to decide whether to proceed with development. Each voter is a point of failure at which this information could leak to a bad actor. A decentralized structure may therefore result in harm being done that could be avoided if the information is only known to a small, centralized group, and thus does not leak.

Furthermore, large groups are not guaranteed to coordinate in ways that result in globally optimal outcomes. Decentralizing AI to shift governance into the hands of many is akin to a modern democracy, and faces similar coordination challenges. Democratic voters are not perfectly informed, nor are their preferences always aligned. Moreover, as the number of voters grows, so do opportunities for misinformation campaigns by those with entrenched interests in particular outcomes. For these reasons, we observe a failure of large groups to agree and coordinate on problems that have significant negative externalities, even when there are known solutions, such as in the case of climate change \cite{cooney_2010_the}. It will therefore be challenging to achieve AI safety through democratized and decentralized governance, particularly when for many cases, most notably the alignment problem, there is no clear solution \cite{https://doi.org/10.48550/arxiv.2012.07532}.

Regarding AGI risks, DAOs create a low-friction avenue for systems to actively acquire their own financial resources. An ‘AI DAO’, with no humans in the loop, could monetize content it creates on the blockchain and use this funding to acquire more data and computational power \cite{mcconaghy_2022}. Moreover, the network security granted by a blockchain means that an AI DAO deployed on it would be difficult to shut down. This point may be moot, however, as it not clear that we could shut down an AGI, even if it was created on centralized servers \cite{Everitt2018AGI}.

In summary, decentralized governance in its current form is not a perfect solution to AI safety. However, as discussed, current centralized forms of governance also have significant limitations. Given the potential for DAOs to formalize independent and diverse work on AI, we recommend further experiments and research to improve DAO implementations and use-cases.

\subsection{Poor Accessibility}
Decentralized AI has the potential for greater accessibility than centralized AI, but this is not guaranteed. In fact, it is at risk of being (and arguably currently is) much less accessible. Part of the challenge of aligning AI with more diverse moral views is that it is an extremely complex discipline at the intersection of computer science and mathematics. If only a relatively small group understand the capabilities and risks of a given technology, how can its use be informed by a range of diverse perspectives? Combine AI with the even more niche field of blockchain, and the number of people who understand the technology enough to have informed opinions about it declines rapidly. While no large-scale academic studies have been conducted, early informal surveys suggest that decentralization does not guarantee high levels of diversity, finding that 79\% of DAO participants are men aged between 20 and 40 (n=422) \cite{pagoulatos_2021_thinking}. Therefore, ongoing efforts in education, training, and support are critical for decentralized AI to meet its potential. We also recommend that formal and thorough demographic surveys are undertaken, the results of which may suggest how DAOs can increase their accessibility and thus improve the governance of future R\&D.

\subsection{End-user Security}
The monetization of private data via blockchains creates opportunities for cybercriminals and scammers. Other applications of blockchain technology, such as cryptocurrencies, already suffer from high levels of fraud and manipulation due to a lack of consistent regulation, the ease of anonymity, and low barriers to entry \cite{alexander2020corruption}. The types of fraud committed include Ponzi schemes, fake token sales, ‘pump and dump’ schemes, and token theft \cite{baum-no-date}. Because data, like cryptocurrencies, are also represented as tokens on a blockchain, there is a low switching cost for bad actors to apply existing methods and technologies toward defrauding data rather than currency markets. Moreover, poor security practices may be more likely in the context of blockchains because their immutability and cryptographic-backing are misperceived as unbreakable security features \cite{zetzsche_2017_the}. As work on decentralized AI continues, and blockchains are increasingly leveraged to store data and manage its access, users must be educated in security threats and best-practices.

\subsection{Data-driven Discrimination}
Some have argued that the ‘datafication’ and ‘monetization’ of personal data is unethical. Indeed, the trade of personal data is arguably prohibited under the current EU Charter of Fundamental Rights and the General Data Protection Regulation (GDPR) \cite{custers_2022_priceless}. Critically, monetization may encourage “vulnerable lower-income people to pour more personal information into an industry that exploits and discriminates against them” \cite{jeong_2019_opinion}. Relatedly, there is already evidence that analytics on big data leads to entrenched biases and discrimination, for example, against the poor \cite{Madden2017PrivacyPA}. As blockchain technology brings transparency to financial data and decision-making, this risk is heightened. In the development of systems that leverage private data, it is therefore critical that data remains anonymous, and potentially discriminatory variables are effectively controlled \cite{articleCalmon}. Existing AI techniques that aim to protect privacy, such as on federated learning \cite{li_2020_federated},  and identify biases, such as explainable artificial intelligence \cite{8466590}, should also be prioritized in future efforts to build safe and equitable decentralized AI systems.

\section{Conclusion}
AI safety is a multi-faceted problem, ranging from data biases to the challenge of building AGI that maximizes human flourishing. Given it is impractical to halt progress in AI, and doing so could forgo its many potential benefits, innovative solutions are required. Rather than relying on small, homogenous groups, we argue that input and decision-making from diverse communities increases the probability of finding such solutions whose benefits are equitably distributed. We propose that blockchain technology is used to facilitate and formalize existing efforts to conduct distributed and democratized work on AI safety. In addition, we highlight the advantages of decentralization in creating human-centered AI, securely accessing greater datasets, and improving auditability. Finally, we identify a number of unique challenges faced by decentralized AI, including the vulnerabilities and accessibility of DAOs, end-user security, and the risks of monetizing personal data.

\section*{Acknowledgments}
This work was supported by a grant from Algovera.ai.

%Bibliography
\bibliographystyle{IEEEtran}
\bibliography{references}

\end{document}